\documentclass[preprint]{revtex4}

\usepackage{graphicx}
\newcommand{\eg}{{\it e.g.}}

\newcommand{\CP}{{\em CP}}
\begin{document}
\bibliographystyle{revtex}

\preprint{IIT-HEP-01/8}

\title{Prospects
for Charm CP Violation Studies
in BTeV\footnote{Presented at the {\sl Snowmass Summer Study on the Future of Particle Physics}, Snowmass, Colorado, June 30--July 21, 2001.}
}



\author{Daniel M. Kaplan}
\email[]{kaplan@fnal.gov}
\affiliation{Illinois Institute of Technology, Chicago, IL 60616}

\author{for the BTeV Collaboration}
\noaffiliation


\begin{abstract}
The BTeV experiment  at Fermilab could 
reconstruct $>$10$^9$ charm decays, three orders of magnitude beyond the 
largest extant sample. The experiment is likely to run during Tevatron 
Run II. It will 
have significant new-physics reach in the areas of charm {\em CP} 
violation, flavor-changing neutral-current and lepton-number-violating decays, 
and $D^0\overline {D^0}$
mixing, and could observe direct {\em CP} violation in Cabibbo-suppressed 
$D$
decays if it occurs at the level predicted by the Standard Model.
\end{abstract}

\maketitle


\section{Introduction}
Particle physics at the turn of the millennium faces two
key  mysteries: the origin of mass and the existence of multiple fermion 
generations. While the former mystery may be resolved
by the LHC, the latter appears to originate at higher mass scales, which can
be studied only indirectly. Such effects as \CP\ violation, mixing, and
flavor-changing neutral or lepton-number-violating currents may hold the key to
physics at these new scales~\cite{Hewett,Pakvasa,Sokoloff}.
Because in the charm sector the Standard Model (SM) contributions to these effects are small, these are areas in which charm studies can provide unique information.
In contrast, in the $s$- and $b$-quark sectors, in which such studies are
typically pursued, there are large SM contributions to mixing and \CP\ 
violation~\cite{Rosner,BCP}, which for new-physics searches constitute backgrounds.

\section{Charm \CP\ Violation}

\begin{figure}[t]
\vspace{-0.25in}
\hspace{-0.25in}
\centerline{\scalebox{.85}{\includegraphics{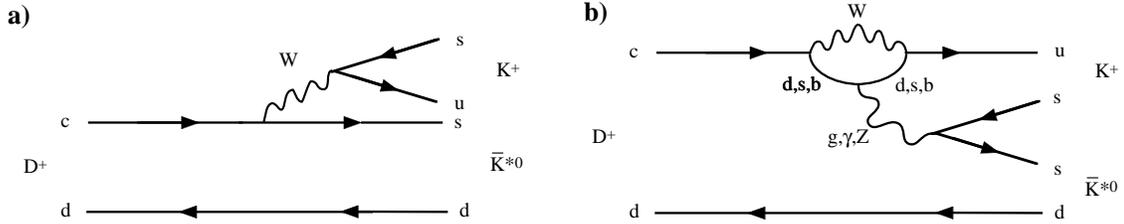}}}
\caption{Example of Cabibbo-suppressed $D^+$ decay that can proceed via  both
(a) tree and (b) penguin diagrams. \label{fig:penguin}
}
\end{figure}

Both direct and indirect \CP\ violation are possible in charm decay. The Standard Model (SM) predicts direct \CP\ violation in singly Cabibbo-suppressed (SCS) charm
decays at the ${\cal
O}(10^{-3})$ level~\cite{charmCP}, arising from the interference of tree-level processes with
penguins (Fig.~\ref{fig:penguin}). The observation of \CP\ asymmetries
substantially larger than this would be unambiguous evidence for new physics, as would almost any
observation of direct \CP\ violation in Cabibbo-favored (CF) or doubly
Cabibbo-suppressed (DCS) charm decays.  The experimental signature for direct \CP\ violation is a difference in partial decay rates between particle and antiparticle:
\begin{equation}
A\equiv\frac{\Gamma(D\to f)-\Gamma(\overline{D}\to \bar f)}{\Gamma(D\to f)+\Gamma(\overline{D}\to \bar f)}\ne 0\,,\label{eq:direct}
\end{equation}
where $f$ and $\bar f$ are \CP-conjugate final states or (for \CP\ eigenstates) $f=\bar f$. In the latter case the two processes of Eq.~\ref{eq:direct} are distinguished by initial-state tagging ($D^{*\pm}\to {}^{^(}\overline{D}{}^{^)}{}^0\pi^\pm$), while in the former case the final states are ``self-tagging." Rates observed in the self-tagging modes typically need to be corrected for production-rate or detection-efficiency asymmetries between particle and antiparticle, hence what is reported is usually an asymmetry normalized to that in a CF mode.

Indirect charm \CP\ violation can arise through the interference of DCS and
mixing amplitudes, \eg\ (for ``wrong-sign" decay of the $D^0$),
\begin{eqnarray}
\Gamma(D^0\to K^+\pi^-)&=&|B|^2\left|\frac{q}{p}\right|^2\frac{e^{-\Gamma t}}{4}
\{4|\lambda|^2+(\Delta M^2+\frac{\Delta\Gamma^2}{4})t^2+[2{\rm Re}(\lambda)\Delta\Gamma +4{\rm Im}(\lambda)\Delta M]t\}\,.\label{eq:mixing}
\end{eqnarray}
(Here we use the notation of Refs.~\cite{Blaylock} and \cite{Browder}.) In
Eq.~\ref{eq:mixing} the first term on the right-hand side is the DCS
contribution, which peaks at $t=0$; the second is the mixing contribution,
which peaks at two $D^0$ lifetimes due to the factor $t^2$; and the third term
reflects interference between mixing and DCS decay and peaks at one lifetime
due to the factor $t$. Given the small values of $\Delta M$ and $\Delta \Gamma$
for the $D^0$~\cite{PDG}, the interference term (which is linear in $\Delta M$
and $\Delta \Gamma$) may be more easily detectable than the pure mixing  term.
\CP\ is violated if $\lambda\ne{\bar \lambda}$, where $\lambda$ (${\bar
\lambda}$) is the DCS amplitude for  $D^0$ ($\overline{D}{}^0$).

A variety of extensions of the Standard Model have been considered~\cite{Nir} 
in which  charm \CP\ asymmetries could be as large as ${\cal O}(10^{-2})$.
These include models with leptoquarks~\cite{Lepto}, extra Higgs doublets (\eg\
non-minimal supersymmetry~\cite{Bigi94}), a fourth
generation~\cite{Pakvasa,Babu}, or  right-handed weak
currents~\cite{Pakvasa,Yaouanc}. In addition, two Standard Model possibilities
for large \CP\ asymmetries in charm have been discussed:  asymmetries due to
$K^0$ mixing in \eg\ $D^\pm\to K_S\pi^\pm$~\cite{Xing}, and  the intriguing
possibility that $D$ mesons mix with glueballs or gluonic 
hybrids~\cite{Close-Lipkin}.

Recent experimental hints~\cite{Godang,Link,Cronin-Hennessy} that $D^0$ mixing
may be on the verge of detection have led to renewed interest in this
physics~\cite{mixing-interest,large-delta}. The experimental situation is that both CLEO
and FOCUS have observed effects at the $\approx$2$\,\sigma$ level:
CLEO measures $y^\prime=(-2.5^{+1.4}_{-1.6}\pm0.3)$\% (for their most general
fit)~\cite{Godang}, where $y^\prime\equiv y\cos{\delta}-x\sin{\delta}$,
$x\equiv\Delta M/\Gamma$, $y\equiv\Delta \Gamma/2\Gamma$, and $\delta$ is the
strong phase between the DCS and CF amplitudes; and FOCUS
obtains~\cite{Link}
$y=(3.42\pm1.39\pm0.74)$\%. (These results, while superficially
contradictory, can be reconciled for a range of possible choices of $\delta$;
whether such large $\delta$ is theoretically plausible is a subject of current
debate~\cite{large-delta}.)

\section{Experimental prospects}

Current sensitivities to charm \CP\ violation are summarized in
Table~\ref{tab:CPlimits}. These are based on samples of up to $\sim10^6$
reconstructed charm decays obtained in the CLEO and FOCUS experiments. The $B$
factories can be expected to obtain samples an order of magnitude larger than
these, which should push \CP-violation limits down to the one-to-few-\% level,
and the COMPASS experiment at CERN may also be competitive at this
level~\cite{COMPASS}.

\begin{table} \caption{World-average charm \CP\ asymmetries  (from
Ref.~\protect\cite{PDG}).} 
\vspace{3mm}
\label{tab:CPlimits} 
\begin{center} 
\begin{tabular}{|ccc|} 
\hline Particle & Mode & Asymmetry \\ 
\hline $D^\pm$ & $K^+K^-\pi^\pm$ & $-0.017\pm0.027$ \\ 
& $K^\pm K^{*0}$ & $-0.02\pm0.05$ \\ 
& $\phi\pi^\pm$ & $-0.014\pm0.033$ \\ 
& $\pi^+\pi^-\pi^\pm$& $-0.02\pm0.04$ \\ \hline
${}^{{}^(}\overline{D}{}^{{}^)}{}^0$ & $K^+K^-$ & $0.026\pm0.035$ \\ 
& $\pi^+\pi^-$ & $-0.05\pm0.08$ \\ 
& $K_S\phi$ & $-0.03\pm0.09$ \\ 
& $K_S\pi^0$ & $-0.018\pm0.030$ \\  
& $K^\pm\pi^\mp$ & $0.02\pm0.20$ \\ \hline 
\end{tabular}
\end{center} 
\end{table}

Other experiments that could potentially extend charm \CP-violation sensitivity
include HERA-$B$, LHC$b$, and BTeV. HERA-$B$ has operated so far in a mode in
which charm decays are efficiently rejected by the trigger, and a significant
upgrade of their DAQ bandwidth would be required in order to record a
substantial sample of hadronic charm decays.  Similarly, LHC$b$ is designed to
reject charm at trigger level in order to concentrate on beauty.

\begin{figure}
\centerline{\rotatebox{90}{\scalebox{0.4}{\includegraphics[bb=11 65 541 721,clip]{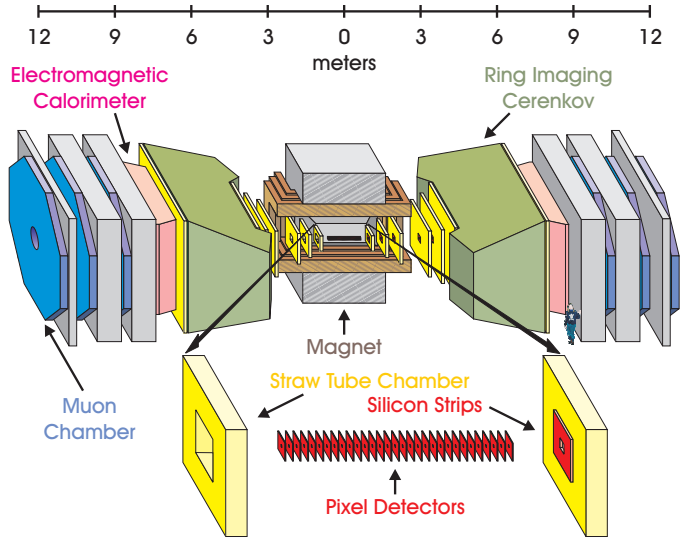}}}}
\caption{Layout of BTeV spectrometer; note that the pixel vertex detector is located within the dipole spectrometer magnet centered on the interaction region.}
\label{fig:BTeV}
\end{figure}

BTeV is an approved Fermilab experiment that will use the Tevatron Collider to
study heavy-quark physics. BTeV data-taking is planned to commence in Tevatron
Run IIB (starting about 2006). To achieve the highest possible sensitivity to
mixing, {\em CP} violation, and rare decays in charm as well as beauty, a very
ambitious trigger system is envisioned that will examine {\em every} beam
crossing for evidence of secondary vertices~\cite{Kaplan-BTeVtrig}.  To do this
it will make use of information from a unique, high-speed, silicon-pixel vertex
detector~\cite{BTeV-proposal,FPIX2} located inside the dipole spectrometer 
magnet (see Fig.~\ref{fig:BTeV}). The design goal for this trigger is an efficiency
better than 50\% for $B$ events of interest.

\begin{figure}
\vspace{0.1in}\centerline{\scalebox{.75}{\includegraphics[bb=0 500 541 750,clip]
{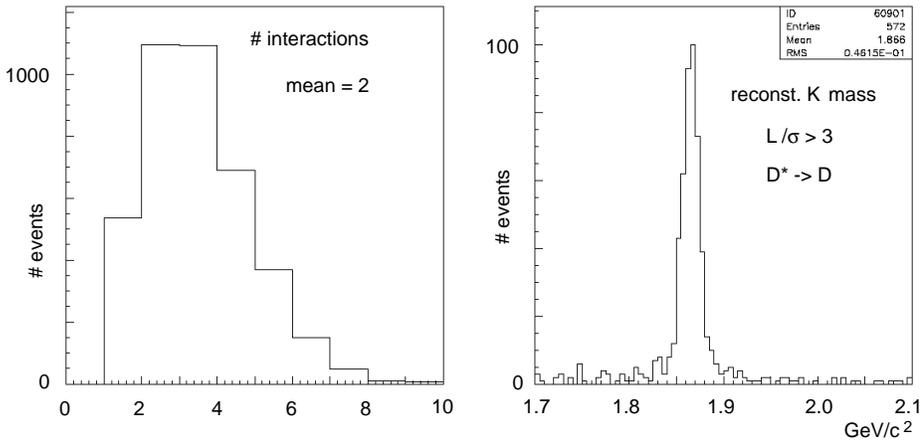}}}
\caption{Preliminary results from a Geant simulation of 4000 $D^*$ decays in 
BTeV~\protect\cite{Butler-Cheung}.}
\label{fig:charmfig}
\end{figure}

Triggering on decay vertices is more difficult for charm than for beauty, since
charm lifetimes are shorter and the typical transverse momenta of charm decay
products (important for vertex resolution) are lower. Fig.~\ref{fig:charmfig}
shows the results of a Geant simulation of the BTeV apparatus, indicating about
1\% efficiency for the decay chain $D^{*+}\to D^0\to K^-\pi^+$, including
geometric acceptance, trigger efficiency, and offline reconstruction
efficiency. (The vertex trigger efficiency is here assumed to be about 10\%,
based on previous simulation results~\cite{Adelaide}, but this needs to be rechecked in
light of the recent evolution of our vertex trigger
algorithm~\cite{Kaplan-BTeVtrig}.) 

Since \CP-violation sensitivity depends in complicated ways on reconstruction
and particle-ID efficiency for various modes, optimization of vertex cuts,
$D^*$-tagging efficiency (for $D^0$ modes), etc., we use here simple overall
benchmarks rather than detailed estimates. These are the total number of charm
decays produced or reconstructed and the total number of
${}^{{}^(}\overline{D}{}^{{}^)}\!{}^0\to K^\mp\pi^\pm$ decays produced or
reconstructed. We scale from current experiments according to the square root
of one of these benchmark numbers to obtain an estimated \CP\ reach,
recognizing that this procedure is at best approximate and addresses only the
statistical component of \CP\ sensitivity. These estimates suggest that BTeV is likely
to reconstruct about $10^8$ such events per year, or $>10^9$-event overall charm
sensitivity summing over all modes. 

Future simulation work should allow
more detailed BTeV charm sensitivity estimates, including both statistical and
systematic effects. However, at the present level of understanding,
sensitivity below the $10^{-3}$ level looks possible in BTeV, thus even charm
\CP\ asymmetries at the Standard Model level may be detectable.

BTeV running might be staged, with wire-target running (\'a l\`a HERA-$B$) in the
years before a high-luminosity $\bar{p}p$ interaction region can be provided
for us. Although collider running is required for competitive beauty
sensitivity, charm sensitivity may be comparable in fixed-target and collider
modes.

\section{Conclusions}

Unique among near-term heavy-quark hadroproduction experiments, BTeV
will have substantial and broad sensitivity to charm events and should be the
leading charm experiment in the latter half of this decade. Sensitivity to
charm \CP\ violation even at the Standard Model level is possible. Observation
of a larger effect could be important evidence for new physics beyond the
Standard Model.


\end{document}